\documentclass[%
reprint,
superscriptaddress,
 amsmath,amssymb,
 aps,
]{revtex4-1}

\usepackage{graphicx}
\usepackage{dcolumn}
\usepackage{bm}
\usepackage[colorlinks=true, 
linkcolor=blue, 
citecolor=blue, 
urlcolor=blue]{hyperref}
\usepackage[normalem]{ulem}
\usepackage{xcolor}

\newcommand{\correct}[2]{%
  \if\relax\detokenize{#1}\relax
  \else
    \textcolor{blue}{\sout{#1}}%
  \fi
  \if\relax\detokenize{#2}\relax
  \else
    \textcolor{red}{%
      \if\relax\detokenize{#1}\relax\else\,\fi
      #2%
    }%
  \fi
}
\newcommand{\paren}[1]{\left( #1 \right)}
\newcommand{\dd}{\mathrm{d}}

\begin{document}

\title{Unbiased Data-Driven Determination of the Nuclear Dipole Amplitude in the Color Glass Condensate}

\author{Si-Wei Dai}
\email[]{swdai@mails.ccnu.edu.cn}
\affiliation{Key Laboratory of Quark and Lepton Physics (MOE) \& Institute of Particle Physics, Central China Normal University, Wuhan 430079, China}
\affiliation{Artificial Intelligence and Computational Physics Research Center, Central China Normal University, Wuhan 430079, China}

\author{Haowu Duan}
\email[]{haowu.duan@ccnu.edu.cn} 
\affiliation{Key Laboratory of Quark and Lepton Physics (MOE) \& Institute of Particle Physics, Central China Normal University, Wuhan 430079, China}
\affiliation{Artificial Intelligence and Computational Physics Research Center, Central China Normal University, Wuhan 430079, China}

\author{Long-Gang Pang}
\email[]{lgpang@ccnu.edu.cn}
\affiliation{Key Laboratory of Quark and Lepton Physics (MOE) \& Institute of Particle Physics, Central China Normal University, Wuhan 430079, China}
\affiliation{Artificial Intelligence and Computational Physics Research Center, Central China Normal University, Wuhan 430079, China}

\author{Guang-You Qin}
\email[]{guangyou.qin@ccnu.edu.cn}
\affiliation{Key Laboratory of Quark and Lepton Physics (MOE) \& Institute of Particle Physics, Central China Normal University, Wuhan 430079, China}
\affiliation{Artificial Intelligence and Computational Physics Research Center, Central China Normal University, Wuhan 430079, China}

\author{Shu-Yi Wei}
\email[]{shuyi@sdu.edu.cn}
\affiliation{Institute of Frontier and Interdisciplinary Science,
Key Laboratory of Particle Physics and Particle Irradiation (MOE), Shandong University, Qingdao, Shandong 266237, China}

\author{Han-Zhong Zhang}
\email[]{zhanghz@ccnu.edu.cn}
\affiliation{Key Laboratory of Quark and Lepton Physics (MOE) \& Institute of Particle Physics, Central China Normal University, Wuhan 430079, China}
\affiliation{Artificial Intelligence and Computational Physics Research Center, Central China Normal University, Wuhan 430079, China}

\author{Wenbin Zhao}
\email[]{WenbinZhao@ccnu.edu.cn}
\affiliation{Key Laboratory of Quark and Lepton Physics (MOE) \& Institute of Particle Physics, Central China Normal University, Wuhan 430079, China}
\affiliation{Artificial Intelligence and Computational Physics Research Center, Central China Normal University, Wuhan 430079, China}

\date{\today}

\begin{abstract}
Gluon saturation limits the growth of parton densities at small Bjorken-$x$ and is expected to be most pronounced in heavy nuclei. Yet quantitative extractions of the nuclear gluon dipole amplitude have long relied on parametrized initial conditions, introducing uncontrolled model dependence that obscures genuine nuclear effects. We introduce a physics-informed neural-network framework that embeds the collinearly improved Balitsky-Kovchegov evolution equation directly into the training objective, allowing the impact-parameter-averaged dipole amplitude to be determined from data without assuming a functional form for its initial condition. Applying this framework to forward-hadron nuclear-modification-factor and coherent $J/\psi$ photoproduction data, we extract the $^{208}$Pb dipole amplitude at $x_0=0.01$ with QCD evolution and momentum-space positivity enforced throughout training. The evolved amplitude reproduces the measured cross sections across the available kinematic range and yields a saturation-scale ratio $Q_{s0,\mathrm{Pb}}^2/Q_{s0,p}^2 = 3.17^{+0.17}_{-0.10}$, consistent with simple geometric scaling. The extracted Pb initial condition is well described by a McLerran-Venugopalan-type form, in contrast to the proton, reflecting the higher color-charge density of a large nucleus. Using the same amplitude, we predict the rapidity dependence of the transverse-momentum ratio in $pp$, $p$Pb, and Pb$p$ collisions, finding agreement with recent LHCb measurements at low multiplicity without any system-dependent parameters. This work provides the first unbiased, data-driven determination of nuclear structure in the saturation regime and establishes a general strategy for embedding nonlinear evolution equations into machine-learning extractions of dynamically constrained observables.
\end{abstract}

\maketitle

\section{Introduction}
At high energy, the internal structure of a proton or nucleus is
overwhelmingly gluonic. As one probes matter at ever smaller values of Bjorken-$x$, linear QCD evolution predicts unbounded growth of the gluon density, which would violate unitarity~\cite{Gribov:1983ivg,Mueller:1985wy}. The resolution is gluon saturation: at sufficiently small $x$, gluon recombination balances gluon splitting, and the system enters a universal, semiclassical regime known as the Color Glass Condensate (CGC)~\cite{Kovchegov:2012mbw,McLerran:2002wj,Iancu:2003xm,Weigert:2005us,Gelis:2010nm,Albacete:2014fwa,Morreale:2021pnn}. This regime is governed by saturation scale $Q_s(x)$ that grows toward small $x$, with $Q_{sA}^2 \propto A^{1/3}$ for nuclei~\cite{McLerran:1993ni,McLerran:1993ka,McLerran:1994vd}. At fixed kinematics, saturation effects are stronger in heavy nuclei than in protons, making nuclei powerful probes of this QCD regime. Determining the onset and quantitative behavior of saturation is essential for understanding hadronic and nuclear structure, constraining heavy-ion initial conditions, interpreting particle production in $e/p+A$ collisions, and modeling ultra-high-energy neutrino and cosmic-ray interactions~\cite{AbdulKhalek:2021gbh,Aschenauer:2017jsk,Accardi:2012qut,Bertulani:2005ru,Klein:2019qfb,Anderle:2021wcy,Arleo:2025oos}.

The color dipole amplitude $N(r,x,b)$ gives the scattering probability of a quark--antiquark dipole of transverse size $r$ on a target at impact parameter $b$. Its energy evolution is governed by the nonlinear Balitsky--Kovchegov (BK) equation~\cite{Balitsky:1995ub,Kovchegov:1999yj} or, more generally, by the Jalilian-Marian--Iancu--McLerran--Weigert--Leonidov--Kovner (JIMWLK) hierarchy~\cite{Jalilian-Marian:1996mkd,Jalilian-Marian:1997qno,Jalilian-Marian:1997jhx,Iancu:2001md,Ferreiro:2001qy,Iancu:2001ad,Iancu:2000hn}. The dipole amplitude underlies a broad class of small-$x$ observables, including inclusive structure functions, particle production, diffraction, and vector-meson production. For the proton, global fits based on parametrized initial conditions~\cite{Albacete:2010sy,Mantysaari:2018zdd,Ducloue:2019jmy,Casuga:2025etc,Casuga:2026xxt,Korcyl:2026nrz} describe HERA $ep$ data well but can violate positivity of the momentum-space dipole amplitude. In previous work, we addressed this limitation using a physics-informed neural network (PINN) that embeds the BK equation in the training and extracts the proton dipole amplitude directly from HERA data without imposing a parametrized initial condition~\cite{Dai:2026nzp}. The resulting amplitude is constrained by QCD evolution, remains positive in momentum space, and avoids bias from a predefined ansatz. This flexibility is particularly important for nuclei, for which existing extractions are far less consistent.

The nuclear dipole amplitude has conventionally been constructed in several ways. For an impact-parameter-averaged amplitude, one assumes the simple scaling $Q_{s0,A}^2 \simeq A^{1/3}Q_{s0,p}^2$, eikonalizes the proton amplitude using an assumed nuclear thickness function, or combines the two. The resulting nuclear initial condition is then evolved with BK/JIMWLK dynamics. Each prescription rests on strong, hard-to-test assumptions about the nuclear profile, with major quantitative consequences. Extracted values of the nuclear initial saturation scale and related quantities vary widely across the literature, and reported values of $Q_{s0,\mathrm{Pb/Au}}^2/Q_{s0,p}^2$ differ by factors of several depending on the underlying assumptions~\cite{McLerran:1993ni,McLerran:1993ka,McLerran:1994vd,Kowalski:2007rw,Deganutti:2023qct,Caucal:2025zkl,Kang:2025vjk}. This uncontrolled model dependence is a major obstacle to using nuclear data as a precision probe of saturation physics and underscores the need for an extraction method that does not impose a functional form on the nuclear initial condition.

A PINN framework offers a natural way to use such data-driven constraints without reintroducing model bias~\cite{RAISSI2019686,9429985,DBLP:journals/corr/abs-2201-05624,Dai:2026nzp,Dai:2026tjl,Kou:2026iau,Baihaqi:2025kjd,Li:2026hnl,Matsuda:2025whj}. The training loss directly encodes the BK evolution equation, together with terms that enforce positivity, the correct asymptotic behavior of the amplitude, and agreement with measured cross sections. The network is thus trained to produce a dipole amplitude that simultaneously obeys QCD evolution, satisfies the physical constraints, and describes the data. Crucially, no a priori parametrization of the  initial condition is imposed at the starting scale $x_0$; instead, it is inferred during training from the evolution dynamics, physical constraints, and experimental data. This eliminates a persistent source of bias in small-$x$ phenomenology, where conclusions about saturation can depend as strongly on the assumed functional form as on the underlying physics. Coherent $J/\psi$ photoproduction in ultraperipheral collisions (UPCs) is particularly well suited to provide such a constraint: a quasi-real photon fluctuates into a $c\bar{c}$ pair that scatters coherently off the target nucleus, and the coherent cross section is proportional to the square of the target gluon dipole amplitude at the relevant dipole size~\cite{Ryskin:1992ui,Iancu:2021rup,Mantysaari:2023qsq}. The nuclear modification factor $R_{p\mathrm{Pb}}$, which compares this process in $p$Pb and $pp$ collisions, further enhances sensitivity to nuclear dynamics. Several otherwise poorly constrained inputs, including the proton PDF, fragmentation functions~\cite{Cassar:2025vdp,Caucal:2025zkl,Fujii:2026ccu}, the choice of the scale in calculations, and higher-order perturbative corrections~\cite{Chirilli:2011km,Chirilli:2012jd,Shi:2021hwx}, enter the $p$Pb and $pp$ cross sections in nearly the same way, so their effects largely cancel in the ratio. Consequently, $R_{p\mathrm{Pb}}$ isolates genuine small-$x$ nuclear physics from the theoretical uncertainties that dominate absolute cross-section predictions, making it a particularly clean training constraint for extracting the Pb dipole amplitude without assuming its initial functional form.

Here we apply this approach to extract the $^{208}$Pb dipole amplitude within the CGC framework, using coherent $J/\psi$ photoproduction cross sections and $R_{p\mathrm{Pb}}$ as the primary experimental constraints. The analysis makes two key advances. First, no parametric form is imposed on the nuclear initial condition; the shape of the amplitude is learned directly from the combined constraints of BK evolution and data. Second, the BK equation is embedded directly in the PINN training objective rather than imposed a posteriori through a fitting template, so the extracted amplitude is constrained to satisfy QCD evolution throughout training. We show that the extracted Pb dipole amplitude reproduces the available coherent $J/\psi$ and $R_{p\mathrm{Pb}}$ data across the measured kinematic range. We then determine the corresponding nuclear saturation scale and map the dependence of the dipole amplitude on nuclear size. We also predict the ratio of transverse momenta as a function of rapidity in $p$-$p$, $p$-Pb and Pb-$p$ collisions and find agreement with the recent LHCb measurement~\cite{LHCb:2026bkz}. The paper is organized as follows: Sec.~II briefly reviews the CGC dipole formalism and the BK evolution equation, then details the PINN architecture and its training procedure; Sec.~III presents the extracted Pb dipole amplitude and its comparison with the $J/\psi$ UPC and $R_{p\mathrm{Pb}}$ data; and Sec.~IV discusses the implications for the broader saturation program at the LHC and the Electron-Ion Collider.

\section{Model and Set-ups}
\subsection{Collinearly improved Balitsky--Kovchegov evolution of the dipole amplitude}
At high energy, the scattering of a quark-antiquark dipole off a hadronic target is characterized by the dipole $S$-matrix
\begin{equation}
    S_F(\boldsymbol r,Y) =
    \frac{1}{N_c}
    \left\langle
    \mathrm{Tr}\,
    V(\boldsymbol x)V^\dagger(\boldsymbol y)
    \right\rangle_Y,
    \qquad
    \boldsymbol r=\boldsymbol x-\boldsymbol y,
\end{equation}
where $V(\boldsymbol x)$ is the fundamental Wilson line at transverse position $\boldsymbol x$, the average $\langle \cdots \rangle_Y$ is taken over target color-field configurations at rapidity $Y = \ln(x_c/x)$, and $x_c = 0.03$ sets the start point for the small-$x$ evolution in our simulation. The corresponding dipole scattering amplitude is $N(r,Y) = 1 - S_F(r,Y)$. Throughout, we assume the target is translationally and rotationally invariant in the transverse plane, so that the dipole amplitude depends only on the dipole size $r = |\boldsymbol r|$ and not on the impact parameter.

With large $N_c$ and mean-field approximation, the Balitsky hierarchy truncates to the Balitsky--Kovchegov equation~\cite{Balitsky:1995ub,Kovchegov:1999yj}. At next-to-leading order, the BK kernel receives collinearly enhanced double-logarithmic corrections that destabilize the evolution, driving the amplitude negative or otherwise unphysical at moderate rapidities~\cite{Iancu:2015vea,Lappi:2016fmu}. We therefore work with the collinearly improved BK (ciBK) equation, which resums both the double and single transverse logarithms to all orders~\cite{Beuf:2014uia,Iancu:2015vea,Iancu:2015joa,Ducloue:2019ezk,Ducloue:2019jmy}:
\begin{widetext}
\begin{align} \label{eq:ciBK}
    \frac{\partial N(r,Y)}{\partial Y}
    ={}&
    \int\frac{\mathrm{d}^2\boldsymbol r_1}{2\pi}\,
    \bar{\alpha}_s^{\mathrm{BLM}}(r,r_1,r_2)\,
    \frac{r^2}{r_1^2\,r_2^2}
    \left[
    \frac{r^2}{\min(r_1^2,r_2^2)}
    \right]^{\!\pm\, \bar{\alpha}_s^{\mathrm{BLM}}(r,r_1,r_2)\,A_1} \notag \\
    &\times \Big[N(r_1,Y-\delta_1)+N(r_2,Y-\delta_2)-N(r,Y)-N(r_1,Y-\delta_1)\,N(r_2,Y-\delta_2)\Big],
\end{align}
\end{widetext}
with $\boldsymbol r_2=\boldsymbol r-\boldsymbol r_1$ and $r_i=|\boldsymbol r_i|$. Here $\bar{\alpha}_s^{\mathrm{BLM}}$ is the running coupling in the Brodsky-Lepage-Mackenzie (BLM) prescription, and $A_1$ is the coefficient of the single transverse logarithm. The rapidity shifts $\delta_i$ equivalently resum the double transverse logarithms, while the power-law prefactor in brackets accounts for the single-logarithmic contribution. The sign in the exponent, the explicit form of the $\delta_i$, and the prescription for the kinematically forbidden region $Y-\delta_i<0$ are specified in Appendix~\ref{app:ciBK} of the Supplemental Material.

Solving Eq.~\eqref{eq:ciBK} repeatedly inside a global fitting loop would be computationally expensive. We therefore represent the dipole amplitude with a physics-informed neural network, constrained by the ciBK residual and by the observables introduced below. This construction provides a smooth, differentiable representation of $N(r,Y)$ throughout the trained $(r,Y)$ domain, at a substantially reduced evaluation cost compared to direct numerical integration of the evolution equation.

\subsection{Observable constraints}
In our framework, two classes of data enter the fit. The first is the nuclear modification factor for forward hadron production in $p$Pb collisions at the LHC, and the second is exclusive $J/\psi$ photoproduction in Pb-Pb ultraperipheral collisions.

\medskip
\noindent\textit{Forward hadron production.}
Forward hadron production in $pA$ collisions constrains the momentum-space dipole distribution through the hybrid factorization formula, after convolution with projectile parton distribution functions (PDFs) and fragmentation functions (FFs). For a target $\mathcal{T} = p$ and Pb, the momentum-space dipole distribution is defined as
\begin{equation}
    \phi_R^{\mathcal{T}}(x,k_T) = \int \dd^2 \boldsymbol{r}\, e^{i \boldsymbol{k}_T \cdot \boldsymbol{r}} \left[1-N_R^{\mathcal{T}}(r,x) \right], \qquad R=F,\;\mathrm{adj}.
\end{equation}
In the leading-order hybrid formalism, the forward hadron spectrum reads~\cite{Dumitru:2005gt,Chirilli:2012jd,Rezaeian:2012ye}
\begin{align}\label{eq:single_hadron}
    \frac{\dd^3\sigma^{p\mathcal{T}\to hX}}{\dd y\, \dd^2p_T}
    ={}& K_h\frac{S_\perp^{\mathcal{T}}}{(2\pi)^2}
    \int_{x_F}^{1}\frac{\dd z}{z^2} \notag \\
    &\times \Bigg[\sum_q x_p f_{q/p}(x_p,\mu^2)\, \phi_F^{\mathcal{T}}(x_g,k_T)\, D_{h/q}(z,\mu^2) \notag \\
    &\quad + x_p f_{g/p}(x_p,\mu^2)\, \phi_{\mathrm{adj}}^{\mathcal{T}}(x_g,k_T)\, D_{h/g}(z,\mu^2) \Bigg],
\end{align}
where $p_T$ is the transverse momentum of produced hadron, $z$ is the fragmentation momentum fraction carried by the hadron, and $k_T = p_T/z$ is the transverse momentum of the parton. Here $x_F = (p_T/\sqrt{s})e^y$ ensures the kinematic constraints $x_p, x_g < 1$, while $x_p = (p_T/(z\sqrt{s}))e^y$ and $x_g = (p_T/(z\sqrt{s}))e^{-y}$ are the longitudinal momentum fractions of the projectile and target, respectively. Following Ref.~\cite{Shi:2021hwx}, we choose the factorization scale $\mu^2 = \alpha^2(\mu_{\min}^2+p_T^2)$ with $\alpha=3$ and $\mu_{\min} = 1~\mathrm{GeV}$ in our main calculations. The factor $K_h$ effectively absorbs missing higher-order corrections and the uncertainties from the fragmentation functions in the single-inclusive hadron spectrum, for both $pp$ and $p$Pb collisions. The adjoint dipole amplitude is:
\begin{equation}
    N_{\mathrm{adj}}^{\mathcal{T}}(r,Y) = 1 - \left[ 1 - N_F^{\mathcal{T}}(r,Y) \right]^{C_A/C_F},
\end{equation}
with $C_A=N_c$ and $C_F=\frac{N_c^2-1}{2N_c}$. The nuclear modification factor is defined as
\begin{equation}
    R_{pA} = \frac{1}{A}\frac{\dd^3 \sigma^{pA} / \dd y\, \dd^2 p_T}{\dd^3 \sigma^{pp} / \dd y\, \dd^2 p_T}.
\end{equation}
Since the same factor $K_h$ enters both the $pp$ baseline and the $pA$ cross section, it cancels in the ratio $R_{pA}$. We therefore choose to fit $R_{p\mathrm{Pb}}$ rather than the single-inclusive hadron spectrum, since the ratio further suppresses higher-order effects~\cite{Shi:2021hwx} and fragmentation-function uncertainties (see Fig.~\ref{fig:R_pPb}). We have also checked that $R_{pPb}$ is only weakly sensitive to the scale parameter $\alpha$, see Appendix~\ref{app:scale} of the Supplemental Material for details.

\medskip
\noindent\textit{Exclusive vector-meson production.}
In the dipole picture, the exclusive vector-meson photoproduction amplitude is~\cite{Kowalski:2006hc}
\begin{equation}
    \mathcal{A}_{p}(x,Q^2) = \int_0^1 \dd z \int_0^{\infty} r\,\dd r\; (\Psi_V^* \Psi_{\gamma})_{\lambda}(r,z,Q^2)\,N(r,x),
\end{equation}
where $\lambda=T,L$ denotes the photon polarization. In this impact-parameter-averaged formulation, the transverse area $S^{\mathcal{T}}_\perp$ is the same one in forward hadron production in Eq.~(\ref{eq:single_hadron}), and kept outside in the Eq.~(\ref{eq:vmcs}). Then the coherent photoproduction cross section is
\begin{equation}\label{eq:vmcs}
    \sigma^{\gamma A \to VA}(W) = K_V \frac{S_{\perp}^{\mathcal{T}}}{4} \sum_{p=T,L} R_g^2(\lambda_{p})(1+\beta_{p}^2)|\mathcal{A}_{p}|^2,
\end{equation}
with $x = M_V^2 / W^2$ for photoproduction, and $M_V = 3.097~\mathrm{GeV}$ is $J/\psi$ mass and $W$ is the photon--nucleon center-of-mass energy. The factor $R_g(\lambda_p)$ accounts for the skewness correction associated with off-forward gluon exchange, while $\beta_p$ is the ratio of the real to imaginary parts of the production amplitude for polarization $\lambda$. The effective exponent $\lambda_{p}$ characterizes the small-$x$ energy dependence of the corresponding amplitude. The factor $K_V$ accounts for the uncertainty associated with the $J/\psi$ wave-function model. Please note that it is fixed to $K_V = 1.079$ as determined in our previous work~\cite{Dai:2026nzp}, not a free parameter in this fitting. The details about the wave-function overlap, skewness and real-part corrections, and normalization conventions are included in Appendix~\ref{app:VM} of the Supplemental Material.

For symmetric ultraperipheral collisions, either nucleus can emit the photon while the other acts as the target. The coherent rapidity distribution is then
\begin{align}
    \frac{\dd \sigma^{AA\to AVA}}{\dd y}
    =& n_A(\omega_+)\,\sigma^{\gamma A\to VA}(W_+) \notag \\
    +& n_A(\omega_-)\,\sigma^{\gamma A\to VA}(W_-),
\end{align}
with $\omega_\pm = (M_V/2)\,e^{\pm y}$, $W_\pm^2 = 2\omega_\pm\sqrt{s_{NN}}$, and $n_A(\omega_{\pm})$ the photon flux evaluated using the STARlight event generator~\cite{Klein:2016yzr}.

We emphasize that in this nuclear dipole-amplitude extraction, $K_h$ cancels in $R_{pA}$, while all other external parameters, such as $K_V$ and the parameters in the ciBK evolution, are fixed from our previous work~\cite{Dai:2026nzp}. The only adjustable nuclear-geometry parameter in this description is the transverse area $S_\perp^{Pb} = \pi R_{Pb}^2$, which enters both the $p\mathrm{Pb}$ forward hadron production cross section in Eq.~(\ref{eq:single_hadron}), and the $J/\psi$ photoproduction cross section of Eq.~(\ref{eq:vmcs}). Since the impact-parameter profile is not resolved, this framework constrains only the average nuclear opacity. A detailed description of the $t$ dependence, diffractive minima, or incoherent vector-meson production is left for future work.

\subsection{Physics-informed neural-network framework}
\begin{figure*}[t]
    \centering
    \includegraphics[width=0.93\textwidth]{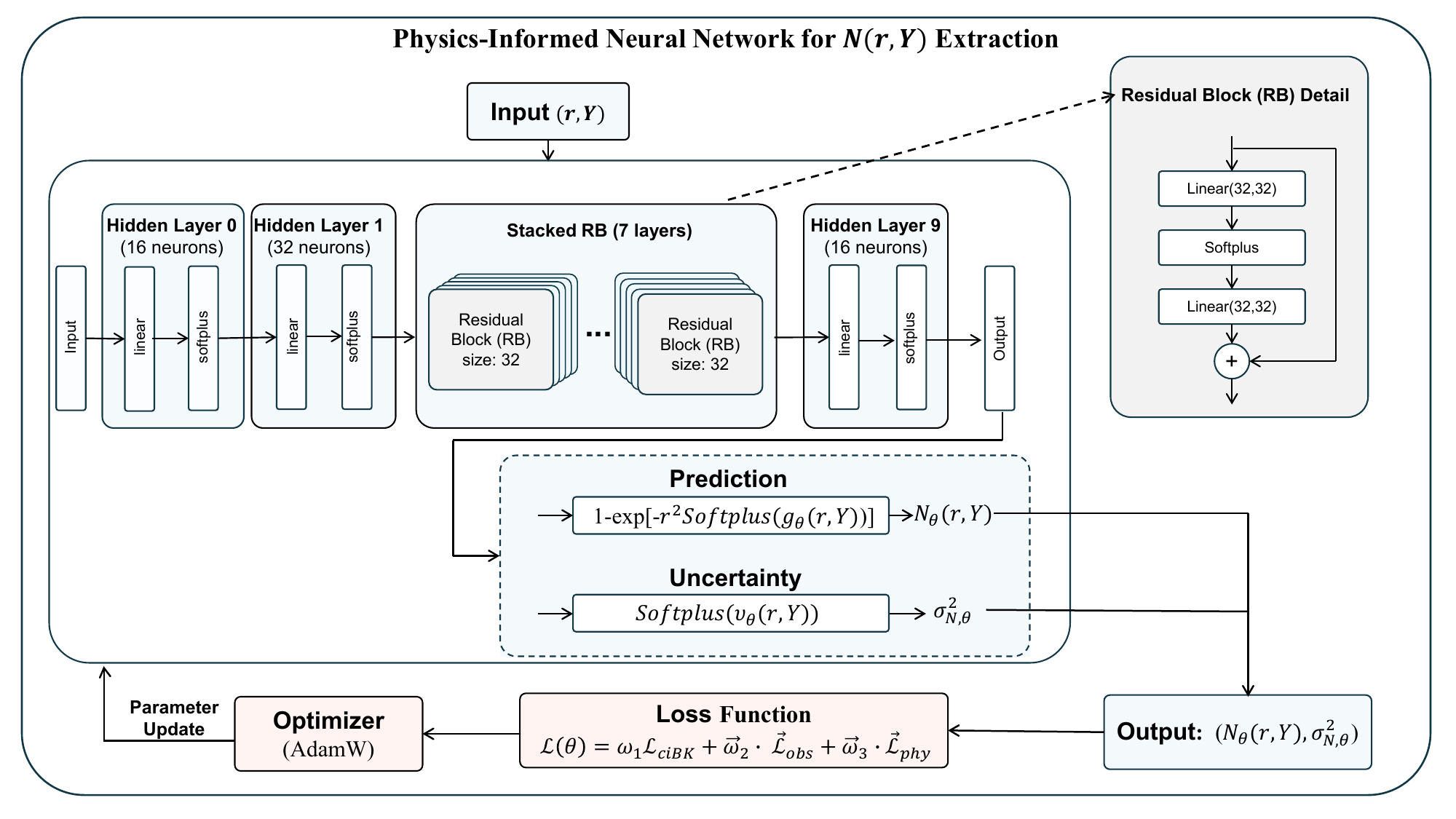}
    \caption{Architecture of the physics-informed neural network used to extract the nuclear dipole amplitude. Given $(r,Y)$ as input, the network outputs the dipole amplitude $N_{\theta}(r,Y)$ and its pointwise predictive variance $\sigma_{N,\theta}^2(r,Y)$. Training is constrained by the ciBK evolution equation, experimental data, and physical boundary conditions.
    }
    \label{fig:PINN_workflow}
\end{figure*}
We formulate the extraction of the nuclear dipole amplitude as a differentiable, physics-constrained inverse problem. The unknown amplitude is represented by a neural network,
\begin{equation}
N(r,Y) \longrightarrow N_{\theta}(r,Y),
\label{eq:NN_representation}
\end{equation}
with $\theta$ denoting the trainable parameters. Rather than solving the ciBK equation repeatedly within the fit, the network is trained to satisfy the ciBK residual while simultaneously reproducing the observable constraints. Once trained, it provides a smooth, differentiable representation of the dipole amplitude over the trained $(r,Y)$ domain at substantially reduced evaluation cost.

As summarized in Fig.\ref{fig:PINN_workflow}, the input $(r,Y)$ is processed by a residual multilayer perceptron and mapped to two scalar outputs, $g_{\theta}(r,Y)$ and $v_{\theta}(r,Y)$. Details of the network architecture, activation functions, and optimization settings are given in Appendix~\ref{app:obs_pro} of the Supplemental Material.

The first output is mapped to the dipole amplitude via
\begin{equation} \label{eq:NN_amplitude_parameterization}
N_{\theta}(r,Y) = 1- \exp \left[
-r^2 \operatorname{Softplus} \bigl(g_{\theta}(r,Y) \bigr) \cdot  Q^2_c \right],
\end{equation}
where $Q_c=1~\mathrm{GeV}$ ensures that the argument of the exponential is dimensionless. Here, $\operatorname{Softplus}$ is a smooth nonlinear function, defined as
\begin{align} \label{eq:softplus}
    \operatorname{Softplus}(z) = {1\over \beta} \log \left(1 + \exp(\beta z) \right),
\end{align}
where the parameter $\beta>0$ controls the smoothness. We set $\beta = 1$ in this work. To avoid numerical overflow, the implementation uses the linear form ${\rm Softplus}(z) \simeq z$ when $\beta\,z > 20$. $\rm Softplus$ is used as the activation function in the hidden layers (see Figure.~\ref{fig:PINN_workflow} and Appendix~\ref{app:obs_pro} of the Supplemental Material), where its smoothness ensures well-defined derivatives for the automatic differentiation entering the ciBK residual and the anomalous-dimension constraint. Its strictly positive output is also used in the parametrizations of both the dipole amplitude and its predictive variance. The amplitude parameterization inherently enforces the physical constraints $0\leq N<1$ and csolor transparency, $N\to 0$ as $r\to 0$. The second output parameterizes the pointwise predictive variance associated with the dipole amplitude $N_\theta(r,Y)$,
\begin{equation} \label{eq:NN_variance_parameterization}
\sigma_{N,\theta}^{2}(r,Y) = \operatorname{Softplus} \bigl(v_{\theta}(r,Y)\bigr) + \epsilon_{\sigma},
\end{equation}
where $\epsilon_{\sigma}>0$ is a small regulator ensuring a strictly positive variance. The learned quantity $\sigma_{N,\theta}^{2}(r,Y)$ characterizes the heteroscedastic uncertainty the network assigns to the dipole amplitude at each point in the $(r,Y)$ domain, and is subsequently propagated to the predicted observables.

\medskip
\noindent\textit{Physics-informed objective.}
The network is trained by minimizing
\begin{align} \label{eq:combined_loss}
    \mathcal L_{\mathrm{tot}}
    ={}&
    w_{\mathrm{ciBK}}\mathcal L_{\mathrm{ciBK}}
    +
    w_{R_{pA}}\mathcal L_{R_{pA}}
    +
    w_{\mathrm{VM}}\mathcal L_{\mathrm{VM}} \notag \\
    &+
    w_{\mathrm{black}}\mathcal L_{\mathrm{black}}
    +
    w_{\mathrm{pos}}\mathcal L_{\mathrm{pos}}
    +
    w_{\gamma}\mathcal L_{\gamma},
\end{align}
where $\mathcal L_{\mathrm{ciBK}}$ penalizes violations of the ciBK evolution equation, $\mathcal L_{R_{pA}}$ constrains the predicted nuclear modification factors against the forward-hadron data, and $\mathcal L_{\mathrm{VM}}$ encodes the constraints from the exclusive $J/\psi$ observables. The remaining three terms enforce physical boundary conditions: $\mathcal L_{\mathrm{black}}$ imposes the large-$r$ saturation (black-disk) limit, $\mathcal L_{\mathrm{pos}}$ enforces momentum-space positivity on the sampled $k_T$ grid, and $\mathcal L_{\gamma}$ constrains the effective anomalous dimension to its prescribed ultraviolet range.

Because the different loss components generate gradients with substantially different magnitudes, their relative weights are determined through a preliminary gradient-calibration stage. The calibration procedure and the fixed numerical weights used in the production runs are reported in Appendix~\ref{app:obs_pro} of the Supplemental Material. Optimization is performed with AdamW~\cite{loshchilov2018decoupled} and a cosine-annealed learning-rate schedule~\cite{DBLP:journals/corr/LoshchilovH16a}.

\medskip
\noindent\textit{ciBK evolution loss.}   
The evolution loss penalizes the ciBK residual,
\begin{equation} \label{eq:ciBK_NN}
    \mathcal R_{\theta}(r,Y) = \frac{\partial N_{\theta}}{\partial Y} -
    \mathcal B_{\mathrm{ciBK}}
    \left[
        N_{\theta}
    \right](r,Y),
\end{equation}
where $\mathcal B_{\mathrm{ciBK}}$ denotes the right-hand side of Eq.~\eqref{eq:ciBK}, including the same running-coupling, rapidity-shift, and boundary prescriptions specified in Appendix~\ref{app:ciBK} of the Supplemental Material. The residual is evaluated over collocation points $\{(r_j, Y_j)\}$, with the sampling strategy detailed in Appendix~\ref{app:obs_pro} of the Supplemental Material, using automatic differentiation for the rapidity derivative and differentiable quadrature for the transverse integral. The corresponding loss function reads
\begin{equation} \label{eq:ciBK_loss}
    \mathcal L_{\mathrm{ciBK}}
    =
    \frac{1}{N_{\mathrm{col}}}
    \sum_{j=1}^{N_{\mathrm{col}}}
    \left|
        \mathcal R_{\theta}(r_j,Y_j)
    \right|^2.
\end{equation}

\medskip
\noindent\textit{Observable losses.}
For each observable class $ c\in \{R_{pA},\mathrm{VM} \}$, we use a heteroscedastic Gaussian loss. For the $i$-th data point in observable class $c$, the prediction and its learned uncertainty contribution are obtained from differentiable forward maps,
\begin{equation}
    O_{c,i}^{\mathrm{th}} = \mathcal F_{c,i} \left[ N_{\theta} \right], \qquad \left(\sigma_{c,i}^{\mathrm{NN}}\right)^2 = \mathcal U_{c,i} \left[ N_{\theta}, \sigma_{N,\theta}^{2}\right].
\end{equation}
Here $\mathcal F_{c,i}$ encapsulates the Fourier transforms, fragmentation and wave-function convolutions, phase-space integrations, and nonlinear operations appropriate to the observable, while $\mathcal U_{c,i}$ denotes the corresponding first-order uncertainty-propagation map. The explicit Jacobian construction, including its discretization and the treatment of quadrature weights, is given in Appendix~\ref{app:obs_pro} of the Supplemental Material.

We adopt the diagonal-covariance approximation, and define $V_{c,i} = \left(\sigma_{c,i}^{\mathrm{NN}}\right)^2 + 
\left(\sigma_{c,i}^{\mathrm{exp}} \right)^2$, the observable loss reads
\begin{align}
    \mathcal L_{\mathrm{obs}} = \sum_{c} \frac{1}{N_{c}}
    \sum_{i=1}^{N_{c}}
    \left[
    \frac{
        \left(
            \mathcal O_{c,i}^{\mathrm{th}}
            -
            \mathcal O_{c,i}^{\mathrm{exp}}
        \right)^2
    }{
        2V_{c,i}
    }
    +
    \frac{1}{2}
    \ln V_{c,i}
    \right],
    \label{eq:loss_obs}
\end{align}
where $N_{c}$ is the number of data points in class $c$. The logarithmic term penalizes overly large learned uncertainties and prevents the network from trivially reducing the standardized residual by inflating $V_{c,i}$.

\medskip
\noindent\textit{Physical regularization.}
The physical loss enforces three conditions: the black-disk limit at large $r$; Fourier positivity of $S_{\theta}(k_T,Y) = 2\pi\int_0^\infty dr\,r\,J_0(k_T r)\,S_{\theta}(r,Y)$; and the bound $0\leq\gamma_{\mathrm{eff}}\leq 1$ on the anomalous dimension $\gamma_{\mathrm{eff}}=\partial\ln N/\partial\ln r^2$ at small$r$. These are implemented as
\begin{align}
    \mathcal L_{\mathrm{phys}}
    ={}&
    \left\|
        N_{\theta}(r_{\max},Y)-1
    \right\|^2_{r_{\max}=50\,\mathrm{GeV}^{-1}}
    \notag
    \\
    &+
    \left\|
        \min
        \left(
            0,\,
            S_{\theta}(k_T,Y)
        \right)
    \right\|^2
    \notag
    \\
    &+
    \Big[
    \left\|
        \max
        \left(
            0,\,
            \gamma_{\mathrm{eff}}-1
        \right)
    \right\|^2
    \notag
    \\
    &+
    \left\|
        \min
        \left(
            0,\,
            \gamma_{\mathrm{eff}}
        \right)
    \right\|^2
    \Big]_
    {r<10^{-3}\,\mathrm{GeV}^{-1}}.
    \label{eq:loss_phys}
\end{align}

The norm in Eq.~\eqref{eq:loss_phys} denotes an average over the corresponding sampled values of $Y$, $k_T$, and $r$. The Fourier-positivity term is imposed as a numerical regularizer on the finite $k_T$ grid used in the calculation.

\medskip
\noindent\textit{Uncertainty quantification.}
Deep ensembles provide a practical approach to uncertainty estimation without relying on Bayesian inference~\cite{lakshminarayanan2017simplescalablepredictiveuncertainty}. We train $M$ independent copies of the network with different random initializations. Each copy $m$ predicts both a central value $N_{\theta_m}(r,Y)$ and a data-dependent (heteroscedastic) variance $\sigma_{\theta_m}^2(r,Y)$, where the latter captures the intrinsic noise level at each input.

The ensemble mean serves as the central prediction,
\begin{equation}
    \overline N(r,Y) = \frac{1}{M}\sum_{m=1}^{M} N_{\theta_m}(r,Y),
\label{eq:ensemble_mean_N}
\end{equation}
and the total variance is decomposed into two components,
\begin{align}
    \sigma_{N,\mathrm{tot}}^{2}(r,Y)
    &=
    \underbrace{
    \frac{1}{M}\sum_{m=1}^{M}
    \sigma_{N,\theta_m}^{2}(r,Y)
    }_{\text{data-dependent}} \notag \\
    &+ 
    \underbrace{
    \frac{1}{M}\sum_{m=1}^{M}
    \left[
        N_{\theta_m} - \overline N(r,Y)
    \right]^2
    }_{\text{model-dependent}}.
    \label{eq:ensemble_total_variance_N}
\end{align}
The first term averages the per-input variances learned by the ensemble members and quantifies the irreducible uncertainty inherent in the data. The second term measures the disagreement among the ensemble members, providing an empirical proxy for sensitivity to initialization and optimization. Although this decomposition does not yield a Bayesian posterior, it offers a practical and computationally efficient separation between data-driven and model-driven uncertainties. Finally, the resulting uncertainties are propagated to the final physical observables via the differentiable forward maps.

\section{RESULTS}
\subsection{Nuclear modification factor and vector meson photoproduction}
In this subsection, we evaluate the nuclear modification factor $R_{p\mathrm{Pb}}$ for single-inclusive forward hadron production at $\sqrt{s_{NN}}=5.02~\mathrm{TeV}$, as well as coherent exclusive $J/\psi$ photoproduction in Pb-Pb ultraperipheral collisions at the LHC.

\begin{figure}[htbp]
    \centering
    \includegraphics[width=0.5\textwidth]{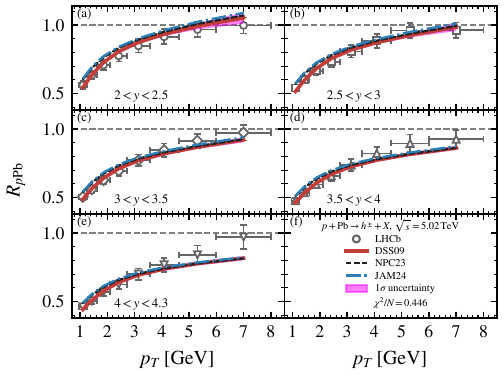}
    \caption{Nuclear modification factor $R_{p\text{Pb}}$ for forward charged-hadron production at $\sqrt{s_{NN}}=5.02~\text{TeV}$, evaluated using three fragmentation-functions, DSS09~\cite{deFlorian:2007aj}, NPC23~\cite{Gao:2024nkz,Gao:2024dbv}, and JAM24~\cite{Anderson:2024evk}. The network is trained with DSS09; the NPC23 and JAM24 results are predictions. Shaded bands denote the $1\sigma$ ensemble uncertainty, and the quoted $\chi^2/\mathrm{d.o.f.}$ uses the solid curve with DSS09 FF. LHCb Data is from Ref.~\cite{LHCb:2021vww}.
    }
    \label{fig:R_pPb}
\end{figure}
We first predict the forward hadron production in $pp$ collisions use our previous PINN extraction~\cite{Dai:2026nzp} of the proton dipole amplitude, and the PDF for dilute part from the CT18NNLO fits~\cite{Hou:2019qau}. As shown in Figure.~\ref{fig:single_hadron_5TeV} of Appendix~\ref{app:single_inclusive} of the Supplemental Material, our extracted amplitude with ciBK evolution predicts the hadron spectra in $pp$ collisions at $5.02~\mathrm{TeV}$ remarkably well. With this baseline established, we compute $R_{p\mathrm{Pb}}$, combined the $J/\Psi$ photoproductions, to globally extract the Pb dipole amplitude.  

Figure~\ref{fig:R_pPb} shows $R_{p\mathrm{Pb}}$ at $\sqrt{s_{NN}}=5.02~\mathrm{TeV}$ evaluated with three different fragmentation-function sets, compared to the LHCb measurements across several rapidity bins. Our PINN extraction reproduces the data well overall, and the agreement is especially good at the relatively low rapidities $2<y<4$. This is consistent with Ref.~\cite{Shi:2021hwx} that found that $R_{p\mathrm{Pb}}$ in this range is largely insensitive to higher-order contributions. At most forward rapidity bin, $4<y<4.3$, our global fit slightly underestimates the $R_{p\mathrm{Pb}}$ data for $p_T \gtrsim 5.5$ GeV. This is the kinematic region where higher-order and threshold resummation effects are expected to become more important~\cite{Shi:2021hwx}, so the discrepancy is not unexpected. Beyond reproducing the data, the ratio also demonstrates a practical advantage. The three curves obtained with the DSS09~\cite{deFlorian:2007aj}, NPC23~\cite{Gao:2024nkz,Gao:2024dbv}, and JAM24~\cite{Anderson:2024evk} fragmentation function sets are nearly indistinguishable, showing that $R_{p\mathrm{Pb}}$ strongly suppresses the uncertainty associated with the choice of non-perturbative fragmentation function. This cancellation is far less effective for the absolute cross sections, which differ by roughly 40\% across the same three FF sets, as shown in Appendix~\ref{app:single_inclusive} of the Supplemental Material.
\begin{figure}[htbp]
    \centering
    \includegraphics[width=0.45\textwidth]{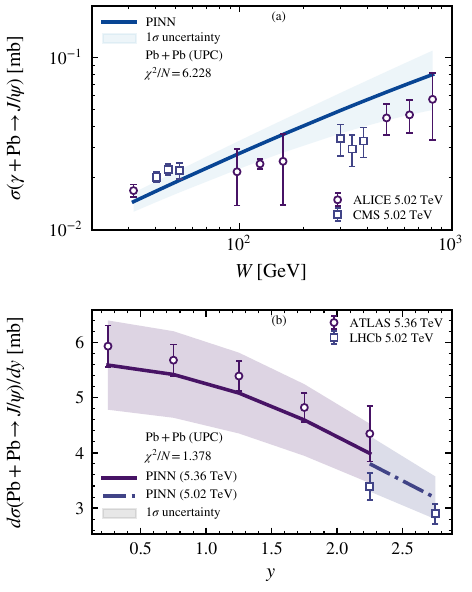}
    \caption{Coherent $J/\psi$ photoproduction in ultraperipheral Pb--Pb collisions. The upper panel shows the coherent photonuclear cross section as a function of the photon--nucleon center-of-mass energy $W$ for $\sqrt{s_{NN}}=5.02~\mathrm{TeV}$, while the lower panel shows the coherent rapidity distributions $\mathrm{d}\sigma/\mathrm{d}y$ at $\sqrt{s_{NN}}=5.02$ and $5.36~\mathrm{TeV}$. The shaded band denotes the $1\sigma$ ensemble uncertainty, and the quoted $\chi^2/\mathrm{d.o.f.}$ uses the solid curve. Data from ALICE~\cite{ALICE:2023jgu}, CMS~\cite{CMS:2023snh}, ATLAS~\cite{ATLAS:2025aav} and LHCb~\cite{LHCb:2022ahs}.
    }
    \label{fig:Jpsi}
\end{figure}

Figure~\ref{fig:Jpsi} shows the coherent $J/\psi$ photoproduction cross section in ultraperipheral Pb--Pb collisions at $\sqrt{s_{NN}}=5.02$ and $5.36~\mathrm{TeV}$, as a function of $W$ and of rapidity, respectively. Compared to the ALICE and CMS data, our global PINN fit slightly underestimates the cross section at low $W$ and overestimates it at high $W$. The rapidity-differential results show a similar pattern, agreeing with the ATLAS data but overestimating the LHCb data. Because our global fit combines $R_{p\mathrm{Pb}}$ with the diffractive $J/\psi$ production data, this discrepancy points to some tension between the UPC $J/\psi$ photoproduction data and the LHCb forward $R_{p\mathrm{Pb}}$ measurement. Part of this tension likely stems from an inherent ambiguity in extracting $\sigma(W)$ from $d\sigma/dy$ in symmetric UPCs. Different event selections, particularly electromagnetic-dissociation tagging with zero-degree calorimeters, probe different impact-parameter ranges and thereby modify the effective photon flux. A more consistent treatment of EMD-tagged fluxes may resolve part of this discrepancy~\cite{Dyndal:2026uvm,Mantysaari:2026vgx}. While the remainder may simply reflect a genuine disagreement between the underlying measurements, most notably the well-known tension between the ALICE and ATLAS results, which our fit cannot fully reconcile. This is consistent with the fact that the uncertainty band on the $J/\psi$ cross section is noticeably broader than that on $R_{p\mathrm{Pb}}$. Since both observable classes are given equal weight in the loss function, this broader band is not an artifact of the fitting procedure but a direct reflection of the larger spread among the current $J/\psi$ measurements themselves.

\subsection{Extracted dipole amplitude}
\begin{figure}[htbp]
    \centering
    \includegraphics[width=0.45\textwidth]{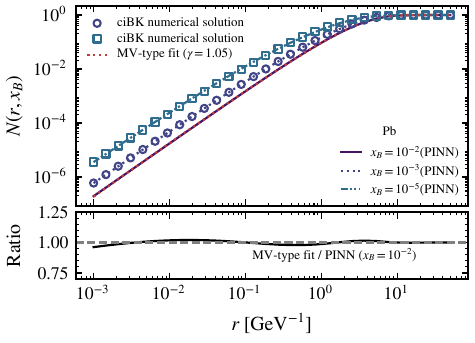}
    \caption{
    Extracted nuclear dipole amplitude $N(r,Y)$ at several values of $x_B$ of Pb. As $x_B$ decreases, the onset of saturation moves to smaller dipole sizes. 
    }
    \label{fig:dipole_r}
\end{figure}

The extracted nuclear dipole amplitude of Pb, obtained from the joint fit to $R_{p\mathrm{Pb}}$ and the coherent $J/\psi$ data with its energy dependence governed by ciBK evolution, is shown in Fig.~\ref{fig:dipole_r}. With this amplitude we also extract an effective Pb radius of $R_{\mathrm{Pb}}=7.83^{+0.15}_{-0.12}~\mathrm{fm}$. As a consistency check, we take the network's initial condition and evolve it independently using a conventional ciBK solver. The resulting solution agrees well with the PINN prediction across the full kinematic domain, confirming that the residual-based training procedure faithfully reproduces the underlying ciBK dynamics. As $x_B$ decreases according to ciBK evolution, the dipole amplitude shifts toward smaller dipole sizes $r$, reflecting the corresponding growth of the saturation scale with decreasing $x_B$.

\begin{figure}[htbp]
    \centering
    \includegraphics[width=0.45\textwidth]{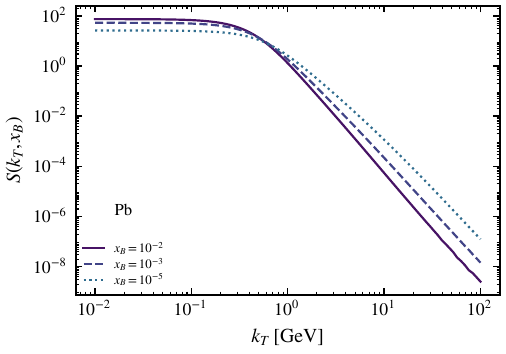}
    \caption{Momentum-space dipole distribution $S(k_T,Y)$ at several values of $x_B$. The distributions are smooth and non-negative across the full kinematic range, confirming that Fourier positivity is preserved under ciBK evolution.
    }
    \label{fig:dipole_k}
\end{figure}

We also use the MV-type parametrization of Eq.~(\ref{eq:MV_model}) to fit the extracted Pb dipole amplitude at $x_0=0.01$, shown in Fig.~\ref{fig:dipole_r} together with the ratio in the lower panel.
\begin{equation} \label{eq:MV_model}
N_{\mathrm{MV}}(r)
=
1-\exp\left[
-\frac{(r^2Q_{s0}^2)^\gamma}{4}
\ln\left(
\frac{1}{r\Lambda}+e_c \cdot \mathrm{e}
\right)
\right].
\end{equation}
It has the best-fit parameters $(Q_{s0}^2,\,\gamma,\,\Lambda,\,e_c) = (0.18~\mathrm{GeV}^2,\,1.05,\,0.11~\mathrm{GeV},\,0.34)$. 
The extracted initial condition of Pb is well described by this MV-type form. 
For comparison, the same MV-type form describes the proton amplitude only to within about 25\%, with deviations that change sign as a function of $r$~\cite{Dai:2026nzp}, whereas it reproduces the Pb amplitude to within about 4\%. This difference is physically expected: for a large nucleus, the high density of color sources makes the Gaussian-source approximation underlying the MV model well motivated~\cite{McLerran:1993ni,McLerran:1993ka,McLerran:1994vd,Jeon:2004rk}, whereas for the proton the assumption $A\gg1$ does not hold, and greater functional flexibility is required to describe the initial condition accurately. We emphasize that this agreement refers to the initial condition at $x_0=0.01$. Under ciBK evolution, the evolved fitted MV-type initial condition deviates from the evolved PINN amplitude at small $x_B$. We also show the Fourier-transformed dipole distribution $S(k_T,Y)$ of Pb in momentum space in Fig.~\ref{fig:dipole_k}. Across the full kinematic range, $S(k_T,Y)$ remains smooth and non-negative, confirming that Fourier positivity is preserved throughout the subsequent ciBK evolution.

At the matching scale $x_0=0.01$, the initial saturation scale is defined via~\cite{Stasto:2000er}
\begin{equation} \label{eq:Qs_definition}
    N\!\left(r=\frac{\sqrt{2}}{Q_{s0}},\,x_B=0.01\right)
    = 1-e^{-1/2}.
\end{equation}
Applying this definition to our unbiased PINN extractions for both the proton and Pb, we obtain the ratio of the impact-parameter-averaged initial saturation scales,
\begin{equation}
    \frac{Q_{s0,\mathrm{Pb}}^2}{Q_{s0,p}^2}
    = 3.17^{+0.17}_{-0.10}.
    \label{eq:Qs_ratio_extracted}
\end{equation}
Since $Q_{s0}^2$ is proportional to the averaged color-charge density, this ratio can be compared to the simple geometric expectation
\begin{equation} \label{eq:Qs_ratio_geometric}
    \frac{Q_{s0,A}^2}{Q_{s0,p}^2}
    \simeq A\,\frac{S_{\perp p}}{S_{\perp A}},
\end{equation}
which gives $2.57^{+0.08}_{-0.10}$ with the PINN-extracted radii $R_p=0.87~\mathrm{fm}$ and $R_{\mathrm{Pb}}=7.83^{+0.15}_{-0.12}~\mathrm{fm}$ for $A=208$. The extracted ratio in Eq.~(\ref{eq:Qs_ratio_extracted}) differs from this geometric estimate by about 20\%, which is well within the accuracy expected of Eq.~\eqref{eq:Qs_ratio_geometric}, since it neglects impact-parameter structure and fluctuations in the color-charge density. Please also note that the saturation-scale definition of Eq.~(\ref{eq:Qs_definition}) yields values slightly larger than those in the MV-type parametrization convention~\cite{Lappi:2013zma}.

\subsection{Ratio of transverse momentum}
With the setup established above from the global fit to $R_{p\mathrm{Pb}}$ and coherent $J/\psi$ photoproduction, we further predict the ratio of the transverse momentum as a function of the laboratory-frame pseudorapidity $\eta_{\mathrm{Lab}}$ for $pp$, $p\mathrm{Pb}$, and $\mathrm{Pb}p$ collisions. This ratio provides a sensitive probe of the interplay between the dynamical evolution of the dipole amplitude, the PDF, and the FF, while largely canceling the uncertainties associated with the FF, the PDF, and higher-order corrections. Moreover, reversing the beam direction interchanges the momentum fractions probed in the projectile and target, offering a nontrivial cross-check of the extracted nuclear initial condition. To compare with the LHCb $p\mathrm{Pb}$ and $\mathrm{Pb}p$ measurements at $5.02~\mathrm{TeV}$, we shift the rapidity by $+0.465$ ($-0.465$) toward the proton-going direction in $p\mathrm{Pb}$ ($\mathrm{Pb}p$) collisions.

\begin{figure}[htbp]
    \centering
    \includegraphics[width=0.45\textwidth]{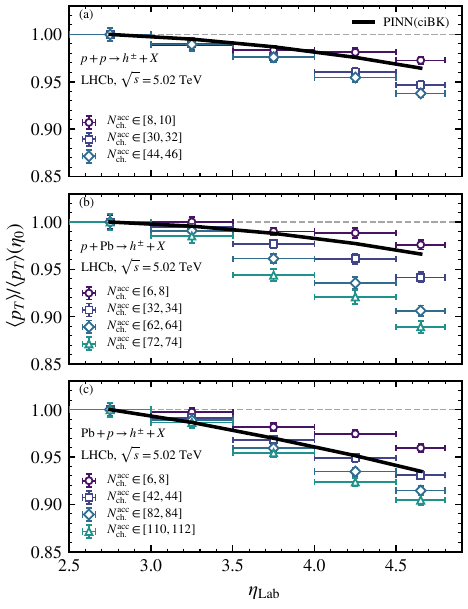}
    \caption{Ratios of the mean transverse momentum
    $\langle p_T\rangle/\langle p_T\rangle(\eta_0=2.75)$ for charged
    hadrons in $pp$ (top), $p\mathrm{Pb}$ (middle), and
    $\mathrm{Pb}p$ (bottom) collisions, compared with 
    LHCb preliminary data~\cite{LHCb:2026bkz} across multiplicity classes.  In our calculation, $\langle p_T \rangle$ is computed using charged hadrons with $p_T > 1$ GeV.
    }
    \label{fig:mean_pt}
\end{figure}

Figure~\ref{fig:mean_pt} compares our predictions with preliminary LHCb data~\cite{LHCb:2026bkz} resolved in multiplicity classes. In $\mathrm{Pb}p$ collisions, the dilute nPDF of the Pb nucleus is described using EPPS21~\cite{Eskola:2021nhw}, dense $proton$ dipole amplitude is from our previous PINN extraction~\cite{Dai:2026nzp}. Our CGC calculation predicts a decreasing trend of the ratio toward forward rapidity. This is  driven by phase-space suppression at large $x$ in the projectile PDF, which outweighs the broadening expected from the growing saturation scale at small $x$ in the target. This trend is reproduced in all three systems without any system-dependent parameters. We note that earlier CGC calculations~\cite{Duraes:2015qoa,Albacete:2012xq} predicted an increasing or flat trend for this ratio. The difference between our results and these earlier calculations likely stems from differences in the assumed initial dipole amplitude and in the details of the BK evolution. Our prediction lies closest to the low-multiplicity data, consistent with a picture in which the ratio is dominated by initial-state dynamics. The growing separation from the data at higher multiplicities likely reflects final-state effects, such as collective expansion and multiple-parton interactions~\cite{Zhao:2020wcd,Noronha:2024dtq,JETSCAPE:2026hdw}, that are not included in the present framework. We do not attempt multiplicity-differential predictions in this work. Mapping the charged-particle multiplicity $N_{\mathrm{ch}}^{\mathrm{acc}}$ onto the saturation scale requires modeling impact-parameter fluctuations, color-charge fluctuations, and detector acceptance effects, all of which we leave to future work.

\subsection{Interpolation to other nuclei}
We further generalize the interaction to an arbitrary nuclear mass number $A$. Specifically, we construct the nuclear dipole amplitude at the initial rapidity corresponding to $x_B=0.01$, and then evolve it in $x_B$ with the collinearly improved BK evolution. In coordinate space we introduce the dipole opacity
\begin{equation}
    \Omega_A(r)\equiv -\ln S_A(r),
\end{equation}
where $S_A(r)$ is the impact-parameter averaged dipole $S$-matrix. In the eikonal approximation, multiple scattering factors multiply at the level of $S_A(r)$, which implies that the corresponding opacities add. We therefore assume
\begin{equation}
    \Omega_A(r,x_B=0.01)=\nu_A\,\Omega_p(r,x_B=0.01),
\end{equation}
where $\nu_A~(\nu_{1}=1)$, is an effective thickness factor.

For the impact-parameter averaged nuclear profile, the relevant gluon density (and hence the effective thickness) is assumed to scale as $A^{1/3}$. This leads to the interpolation prescription~\cite{Deganutti:2023qct}
\begin{equation} \label{eq:nu_A}
    \nu_A = 1 + (\nu_{\mathrm{Pb}}-1)\,
    \frac{A^{1/3}-1}{A_{\mathrm{Pb}}^{1/3}-1}.
\end{equation}
Since $S_A(r)=\exp[-\Omega_A(r)]$, the opacity scaling implies
\begin{equation}
    S_A(r)=\exp[-\nu_A\Omega_p(r)]
    =\Big(\exp[-\Omega_p(r)]\Big)^{\nu_A}
    =S_p(r)^{\nu_A}.
\end{equation}
Using in addition relation, $S_{\mathrm{Pb}}(r)=S_p(r)^{\nu_{\mathrm{Pb}}}$, we obtain
\begin{equation}
    S_A(r)=\Big(S_{\mathrm{Pb}}(r)\Big)^{\lambda_A},
    \qquad
    \lambda_A\equiv \frac{\nu_A}{\nu_{\mathrm{Pb}}}.
\end{equation}
With $N(r)=1-S(r)$, this yields the corresponding dipole amplitude
\begin{equation} \label{eq:N_A}
    N_A(r)=1-\left[1-N_{\mathrm{Pb}}(r)\right]^{\lambda_A}.
\end{equation}
In the dilute limit $N_{\mathrm{Pb}}\ll 1$, Eq.~\eqref{eq:N_A} reduces to $N_A\simeq \lambda_A\,N_{\mathrm{Pb}}$, i.e. linear scaling with the effective nuclear thickness, as expected.

The parameter $\nu_{\mathrm{Pb}}$ is fixed by requiring that the interpolation reproduces the proton saturation scale. Implementing Eq.~\eqref{eq:N_A} at $A=1$ and using the saturation-scale definition $N_p(r=\sqrt{2}/Q_{s0,p},x_B=0.01)=1-e^{-1/2}$,
we obtain:
\begin{equation} \label{eq:nu_Pb}
    \nu_{\mathrm{Pb}}
    =-2\ln\!\left[
    1-N_{\mathrm{Pb}}\!\left(
    \frac{\sqrt{2}}{Q_{s0,p}},\,x_B=0.01
    \right)\right].
\end{equation}

\begin{figure}[htbp]
    \centering
    \includegraphics[width=0.45\textwidth]{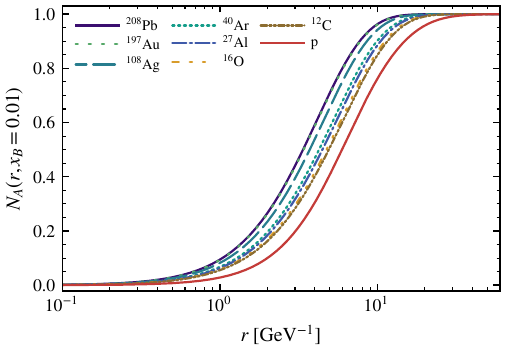}
    \caption{
    The dipole amplitudes $N(r,x_B=0.01)$ for the proton, $^{12}\mathrm{C}, ^{16}\mathrm{O}, ^{27}\mathrm{Al}, ^{40}\mathrm{Ar}, ^{108}\mathrm{Ag}, ^{197}\mathrm{Au}$ and $^{208}\mathrm{Pb}$.
    }
    \label{fig:initial_N_nuclear}
\end{figure}

Figure~\ref{fig:initial_N_nuclear} shows the resulting initial dipole amplitudes
$N(r,x_B=0.01)$ for the proton and for $^{12}\mathrm{C}$, $^{16}\mathrm{O}$, $^{27}\mathrm{Al}$, $^{40}\mathrm{Ar}$, $^{108}\mathrm{Ag}$, $^{197}\mathrm{Au}$, and $^{208}\mathrm{Pb}$. The profiles are smoothly ordered in $A$ (increasing from right to left). Starting from the unbiased, data-driven initial amplitudes for the proton and $^{208}\mathrm{Pb}$, we construct intermediate nuclei through the opacity interpolation described above, and then evolve all systems in $x_B$ using the ciBK equation. These impact-parameter averaged initial conditions therefore provide benchmark inputs for small-$x$ phenomenology at the LHC and RHIC, and for future measurements at the Electron--Ion Collider.

\section{Conclusion}
For the first time, we have presented the data-driven extraction of the impact-parameter-averaged dipole amplitude of $^{208}$Pb within the Color Glass Condensate framework, obtained directly from forward $R_{p\mathrm{Pb}}$ and coherent $J/\psi$ photoproduction  data using a physics-informed neural network that embeds the collinearly improved BK equation in its training objective. No parametric form was assumed for the nuclear initial condition. The shape of the amplitude at the starting rapidity emerged entirely from the interplay of QCD evolution and experimental constraints. This extends to nuclei the strategy we previously developed for the proton~\cite{Dai:2026nzp} and removes a persistent source of model dependence in the nuclear saturation literature, where the assumed initial profile has historically driven much of the spread in extracted saturation scales.

The resulting $^{208}$Pb dipole amplitude, evolved with the ciBK equation, reproduces the measured $R_{p\mathrm{Pb}}$ and coherent $J/\psi$ cross sections across the available kinematic range, confirming that the network faithfully encodes the underlying evolution dynamics while remaining constrained by data. We obtain the saturation-scale ratio $Q_{s0,\mathrm{Pb}}^2/Q_{s0,p}^2 = 3.17^{+0.17}_{-0.10}$ at $x_B=0.01$ and an effective transverse radius $R_{\mathrm{Pb}}=7.83^{+0.15}_{-0.12}~\mathrm{fm}$, in reasonable agreement with the simple geometric estimate $Q_{s0,A}^2/Q_{s0,p}^2\simeq A\,S_{\perp p}/S_{\perp A}$. This provides, for the first time in a model-independent setting, a quantitative test of the accuracy of this widely used geometric scaling argument. A further notable outcome is that the extracted Pb initial condition, unlike the proton's, is well described a posteriori by an MV-type functional form, consistent with the higher color-charge density expected in a large nucleus, which improves the validity of the underlying Gaussian-source approximation. Because this behavior was not imposed but emerged from the fit, it constitutes an internal consistency check on the unparametrized extraction rather than an assumption built into it.

Two additional tests support the robustness and predictive power of the extracted amplitude. First, evaluating $R_{p\mathrm{Pb}}$ with fragmentation-function sets not used in training leaves the result essentially unchanged, confirming that this observable isolates genuine nuclear dipole dynamics from fragmentation uncertainties, as intended; $R_{p\mathrm{Pb}}$ is also likewise largely insensitive to higher-order QCD effects, the fragmentation functions and to the choice of factorization scale. Second, the same amplitude predicts the rapidity dependence of the transverse-momentum ratio in $pp$, $p\mathrm{Pb}$, and $\mathrm{Pb}p$ collisions without introducing any system-dependent parameters, and this prediction agrees with recent LHCb data~\cite{LHCb:2026bkz} in the low-multiplicity regime, where initial-state dynamics are expected to dominate. The growing deviation at higher multiplicities is a natural signature of final-state effects that lie outside the scope of the present dipole-only framework. Finally, we constructed physically motivated initial conditions for nuclei intermediate between the proton and $^{208}$Pb by interpolating the extracted opacity, providing a phenomenological baseline that can be tested directly once additional small-$x$-sensitive observables become available for other nuclear species.

Overall, these results demonstrate that embedding a nonlinear QCD evolution equation directly into the training of a neural network can convert a traditionally model-dependent extraction problem into one governed primarily by data and first-principles dynamics. The strategy is not specific to the dipole amplitude or to gluon saturation: any observable whose evolution obeys a known dynamical equation, and which is constrained by data over only a subset of the physically relevant kinematic range, stands to benefit from the same approach. For the saturation program in particular, an unbiased nuclear dipole amplitude removes a longstanding ambiguity in comparing saturation predictions across nuclear species and provides a firmer foundation for interpreting forward particle production at the LHC and RHIC, as well as future measurements at the Electron-Ion Collider. Extending the present framework to incorporate higher-order corrections and to resolve impact-parameter fluctuations would allow the same data-driven philosophy to be applied to coherent diffractive minima and incoherent vector-meson production, both of which are sensitive to nuclear geometry fluctuations beyond the reach of the impact-parameter-averaged treatment adopted here.

\textbf{\textit{Published Tables}:} The proton, $^{12}\mathrm{C}$, $^{16}\mathrm{O}$, $^{27}\mathrm{Al}$, $^{40}\mathrm{Ar}$, $^{108}\mathrm{Ag}$, $^{197}\mathrm{Au}$, and $^{208}\mathrm{Pb}$ dipole amplitudes shown in this paper, evolved with the ciBK equation, are tabulated together with their momentum-space Fourier transforms in Ref.~\cite{pA_public_table}.

\acknowledgments
\textbf{\textit{Acknowledgments}} We thank  Xiaoxuan Chu, Piotr Tomasz Korcyl, Fu-Peng Li, Wei Li,  Heikki M\"antysaari, Farid Salazar,  and Bj\"orn Schenke  for helpful comments. This work was supported in part by the NSFC under grant No.~12225503, No.~12535010, No.~12435009, No.~12405156, and by the Outstanding Leading Talent Team Program of Central China Normal University (XJ2026000302).  This work was supported also in part by the Cross Research Project of Fundamental Research Funds for Central Universities of Central China Normal University in 2025: ``Advanced Detection and Artificial Intelligence at the Frontiers of Physics'' (No.30101250317).

\clearpage
\onecolumngrid
\appendix

\section*{Supplementary Material for ``Unbiased Data-Driven Determination of the Nuclear Dipole Amplitude in the Color Glass Condensate"}

The Supplemental Material specifies the details of the ciBK evolution, including the running-coupling prescription, rapidity shifts, and boundary treatment. It also details the single-inclusive hadron calculation, scale-variation checks, vector-meson conventions, ultraperipheral collision kinematics, neural-network implementation and uncertainty propagation.

\section{Implementation of the collinearly improved BK evolution}
\label{app:ciBK}
The dipole scattering amplitude $N(r,Y)$ evolves in rapidity according to the Balitsky--Kovchegov (BK) equation, which governs the small-$x$ growth of the gluon density. For phenomenological applications requiring precision agreement with data, we adopt the collinearly-improved BK (ciBK) equation~\cite{Ducloue:2019ezk, Ducloue:2019jmy}, in which exact kinematic constraints are imposed and the associated large collinear logarithms are resummed to all orders. The resulting evolution equation reads
\begin{align} \label{eq:ciBK_form}
\frac{\partial N(r,Y)}{\partial Y} ={}& \int\frac{\mathrm{d}^2\boldsymbol r_1}{2\pi}\,
    \bar{\alpha}_s^{\mathrm{BLM}}(r,r_1,r_2)\,
    \frac{r^2}{r_1^2\,r_2^2}
    \left[
    \frac{r^2}{\min(r_1^2,r_2^2)}
    \right]^{\!\pm\, \bar{\alpha}_s^{\mathrm{BLM}}(r,r_1,r_2)\,A_1}  \notag \\
&\times \mathcal{K}_{\text{DLA}}(\rho) \Big[ N(r_1, Y-\delta_1) + N(r_2, Y-\delta_2) -N(r, Y) - N(r_1, Y-\delta_1)N(r_2, Y-\delta_2) \Big],
\end{align}
where $\mathbf{r}_2 = \mathbf{r}-\mathbf{r}_1$ is the size of the second daughter dipole, and the kernel is built from three multiplicative pieces. 1) The leading-order dipole splitting kernel combined with a running-coupling prescription, 2) the double-logarithmic-approximation (DLA) factor resumming double collinear logarithms, 3) and the single transverse logarithmic (STL) correction. The running coupling itself, $\bar{\alpha}_{\text{BLM}}$, follows the Brodsky--Lepage--Mackenzie (BLM) prescription~\cite{Brodsky:1982gc} and depends on all three dipole sizes involved in the splitting, as given explicitly in Eq.~\eqref{eq:alpha_BLM}. We describe each of these four ingredients, the kinematic constraints, the DLA factor, the STL correction, and the running-coupling prescription, in turn below.

\subsection{ Kinematic constraints (rapidity shifts).}

Consistency of the small-$x$ evolution requires successive soft-gluon emissions to be ordered in lifetime; this constraint is implemented by shifting the rapidity arguments of the daughter-dipole amplitudes appearing in the real-emission terms,
\begin{equation}
\delta_i = \max \{0, \ln \frac{r^2}{r_i^2} \}, \qquad i=1,2,
\end{equation}
where $r$ is the parent-dipole size and $r_1$ and $r_2$ are the sizes of the two daughter dipoles. The shift $\delta_i$ ties the rapidity available to a daughter dipole to its transverse size, such that splittings producing a much smaller daughter dipole are accompanied by a correspondingly larger rapidity shift. This excludes configurations that would otherwise violate lifetime ordering and resums, to all orders, the dominant double collinear logarithms associated with time-ordered emissions, without altering the underlying real--virtual structure of the BK equation.

In evaluating the ciBK residual in Eq.~\eqref{eq:ciBK}, the rapidity shifts render the evolution nonlocal in $Y$, requiring a boundary prescription whenever $Y-\delta_i<Y_0$. In this work, we adopt the frozen-boundary prescription,
\begin{equation}
    N(r,Y<Y_0)=N(r,Y_0),
\end{equation}
where $Y_0$ denotes the starting rapidity of the evolution. In the convention $Y=\ln(x_c/x)$, one has $Y_0=0$.

\subsection{Double logarithmic approximation factor}

An alternative route to resumming the same double collinear logarithms, used in formulations that dispense with explicit rapidity shifts, is to insert the closed-form Bessel factor
\begin{equation}
\mathcal{K}_{\text{DLA}}(\rho) = \frac{J_1(2\sqrt{\bar{\alpha}_s \rho^2})}{\sqrt{\bar{\alpha}_s \rho^2}}  = 1 - \frac{\bar{\alpha}_s \rho^2}{2} + \frac{(\bar{\alpha}_s \rho^2)^2}{12} + \cdots,
\end{equation}
with $\rho = \sqrt{L_{r_1 r}\, L_{r_2 r}}$ and $L_{r_i r} = \ln(r_i^2/r^2)$. Expanding the rapidity-shifted amplitudes $N(r_i, Y-\delta_i)$ order by order in $\bar{\alpha}_s$ generates exactly the same double-logarithmic series $(\bar{\alpha}_s\rho^2)^n$ encoded in $\mathcal{K}_{\text{DLA}}(\rho)$, confirming that the two prescriptions — rapidity shifts and the DLA factor — resum identical physics. Including both simultaneously would therefore double-count the double-logarithmic contributions. Because this resummation is already accounted for by the shifts $\delta_i$, we retain only the leading term of the Bessel expansion, $\mathcal{K}_{\text{DLA}}(\rho)=1$, a choice that keeps the evolution speed stable and avoids spurious oscillations introduced by the higher-order Bessel terms.

\subsection{ Single transverse logarithmic correction}

Besides the double logarithms discussed above, the NLO corrections to the BK kernel contain single transverse logarithms, $\ln(r^2/r_i^2)$, that lie outside the reach of the time-ordering constraints. All orders of this logarithm are resummed through the STL factor
\begin{equation}
\mathcal{K}_{\text{STL}} = \left[ \frac{r^2}{\min(r_1^2, r_2^2)} \right]^{\pm A_1 \bar{\alpha}^{\text{BLM}}}, \qquad A_1 = \frac{11}{12},
\end{equation}
with the coefficient $A_1$ fixed by the one-loop DGLAP anomalous dimension. The exponent's sign depends on the dipole hierarchy: it is positive for $r^2<\min(r_1^2,r_2^2)$ and negative for $r^2>\min(r_1^2,r_2^2)$, ensuring that the kernel is uniformly suppressed in configurations with strongly ordered dipole sizes. Combined with the kinematic constraints of (i) and the truncated DLA factor of (ii), this term completes the collinearly improved kernel used throughout this work.

\subsection{ Running-coupling prescription}

Among the NLO corrections to the BK equation, those enhanced by the one-loop $\beta$-function — the running-coupling corrections — are both sizable and theoretically well established. Several schemes for incorporating them exist in the literature; we adopt the Brodsky--Lepage--Mackenzie (BLM) prescription, also known as ``fast apparent convergence"~\cite{Iancu:2015joa}, in which the coupling entering the kernel, $\bar{\alpha}_{\text{BLM}}(r,r_1,r_2)$, interpolates smoothly between the scales set by all three dipoles participating in the splitting, according to
\begin{equation} \label{eq:alpha_BLM}
\bar{\alpha}_{\text{BLM}}(r, r_1, r_2) = \left[
\frac{1}{\bar{\alpha}_s(r)} + \frac{r_1^2 - r_2^2}{r^2} \frac{\bar{\alpha}_s(r_1) - \bar{\alpha}_s(r_2)}{\bar{\alpha}_s(r_1) \bar{\alpha}_s(r_2)} \right]^{-1},
\end{equation}
where $\bar{\alpha}_s(r)\equiv N_c\,\alpha_s(r)/\pi$. This prescription absorbs the dominant $\beta_0$-dependent NLO contributions directly into the kernel and, in any limit where one dipole is parametrically smaller than the other two, reduces automatically to the natural minimal-dipole coupling $\bar{\alpha}_s(r_{\min})$ — thereby minimizing the impact of the remaining higher-order terms.
The one-loop coupling $\alpha_s(r)$ entering the above expressions is evaluated at the coordinate-space scale $\mu^2=4C^2/r^2$, with the dimensionless constant $C$ — arising from the coordinate-to-momentum-space Fourier transform — treated as a free fit parameter. However, in order to maintain consistency with our previous work~\cite{Dai:2026nzp}, we adopted $C^2 = 0.955$ extracted from our previous work in this paper. Heavy-quark thresholds are included by letting the number of active flavors step from $n_f=3$ to $4$ and then to $5$ as $\mu^2$ crosses $m_c^2$ and $m_b^2$. The infrared Landau pole is regulated by a smooth freezing prescription~\cite{Albacete:2004gw},
\begin{equation}
\alpha_s(r) = \frac{1}{\beta_0^{(n_f)} \ln\left( \frac{4C^2}{r^2 \Lambda_{(n_f)}^2} + \lambda_0 \right)}, \quad \beta_0^{(n_f)} = \frac{11N_c - 2n_f}{12\pi},
\end{equation}
where $\lambda_0 = \exp\!\big[1/(\beta_0^{(3)}\alpha_{\text{fr}})\big]$ guarantees that the coupling approaches a maximal frozen value $\alpha_{\text{fr}}$ deep in the infrared. Continuity across each flavor threshold $\mu^2=m_f^2$ ($f=c,b$) is imposed by recursively rematching the scale parameter,
\begin{equation}
\Lambda_{(n_f-1)} = m_f \left[ \left( \frac{m_f^2}{\Lambda_{(n_f)}^2} + \lambda_0 \right)^{\beta_0^{(n_f)} \over \beta_0^{(n_f-1)}} - \lambda_0 \right]^{-1/2},
\end{equation}
with the recursion anchored at the $Z$-pole value $\Lambda_{(5)}$, fixed from the world-average $\alpha_s(M_Z)$. The fixed parameters used throughout the global extraction are $N_c=3$, $m_c=1.4~\text{GeV}$, $m_b=4.18~\text{GeV}$, $M_Z=91.2~\text{GeV}$, $\alpha_s(M_Z^2)=0.118$~\cite{ParticleDataGroup:2026aaa}, and $\alpha_{\text{fr}}=0.7$~\cite{Albacete:2010sy}.

\section{Single-inclusive hadron production}
\label{app:single_inclusive}
In this section, we present differential cross sections for single-inclusive hadron production in proton--proton ($pp$) and proton--nucleus ($pA$) collisions at RHIC and the LHC. The purpose is twofold: to illustrate the theoretical sensitivity of the absolute hadron spectra to the choice of nonperturbative fragmentation functions (FFs), and to motivate the use of the nuclear modification factor $R_{pA}$ in the main analysis.

\begin{figure}[htbp]
    \centering
    \includegraphics[width=0.45\textwidth]{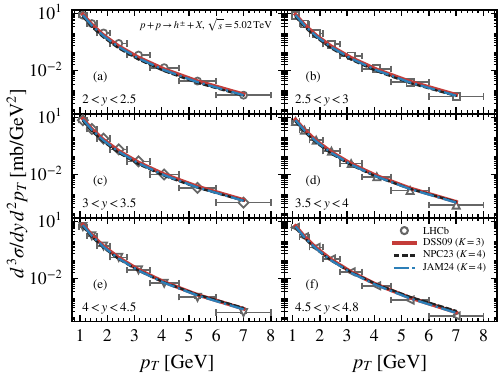}
    \includegraphics[width=0.45\textwidth]{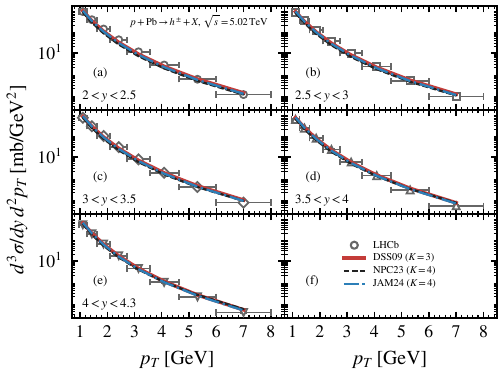}
    \caption{Differential cross sections for forward single-inclusive hadron production in $pp$ (left) and $p\mathrm{Pb}$ (right) collisions at $\sqrt{s_{NN}}=5.02~\mathrm{TeV}$, computed using the DSS09~\cite{deFlorian:2007aj}, NPC23~\cite{Gao:2024nkz,Gao:2024dbv}, and JAM24~\cite{Anderson:2024evk} fragmentation-function sets. The LHCb data is from Ref.~\cite{LHCb:2021vww}.}
    \label{fig:single_hadron_5TeV}
\end{figure}

\begin{figure}[htbp]
    \centering
    \includegraphics[width=0.45\textwidth]{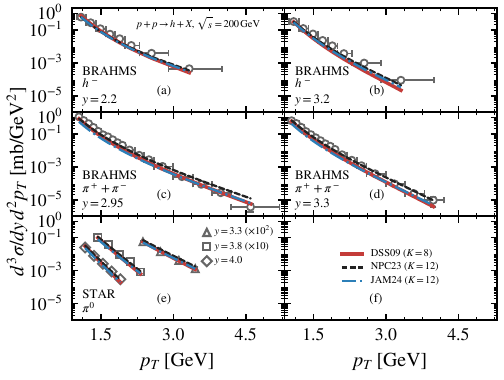}
    \includegraphics[width=0.45\textwidth]{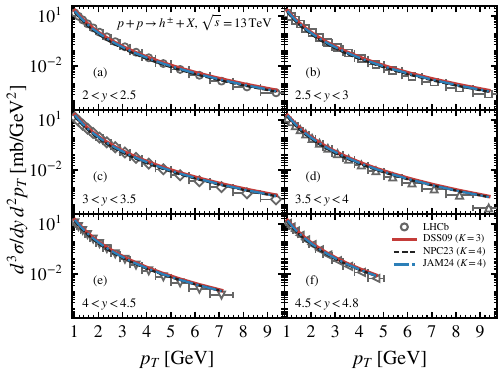}
    \caption{Differential cross sections for forward single-inclusive hadron production in $pp$ collisions at $\sqrt{s}=200~\mathrm{GeV}$ at RHIC (left) and $\sqrt{s}=13~\mathrm{TeV}$ at the LHC (right), computed using the DSS09~\cite{deFlorian:2007aj}, NPC23~\cite{Gao:2024nkz,Gao:2024dbv}, and JAM24~\cite{Anderson:2024evk} fragmentation-function sets. The BRAHMS data is from Ref.~\cite{BRAHMS:2004xry,BRAHMS:2007tyt}, STAR data~\cite{STAR:2006dgg} and LHCb data~\cite{LHCb:2021abm}.}
    \label{fig:pp_rhic}
\end{figure}

We evaluate the hadron spectra using three FF sets: DSS09~\cite{deFlorian:2007aj}, NPC23~\cite{Gao:2024nkz,Gao:2024dbv}, and JAM24~\cite{Anderson:2024evk}. As shown in Figs.~\ref{fig:single_hadron_5TeV}, the three sets yield nearly identical transverse-momentum spectral shapes, while their absolute normalizations differ substantially. To match the experimental cross sections, we introduce an FF-dependent phenomenological factor $K_h$ for each set, which also effectively absorbs normalization corrections associated with missing higher-order perturbative contributions and other theoretical effects not captured at leading order. This pronounced FF dependence introduces a sizable normalization uncertainty into the absolute cross sections. However, at fixed collision energy and for a given fragmentation-function set, $K_h$ is the same in $pp$ and $p\mathrm{Pb}$ collisions, and therefore cancels identically in the ratio $R_{pA}$. This cancellation underlies the robustness of using $R_{pA}$ in the global extraction of the nuclear dipole amplitude presented in the main text. Indeed, the near-overlap of the $R_{pA}$ curves obtained with the different FF sets confirms that $R_{pA}$ is considerably more robust than the absolute hadron spectra, providing a cleaner observable for constraining the initial-state nuclear dipole amplitude.

We further use our extracted dipole amplitude to predict single-inclusive hadron production in $pp$ collisions at RHIC ($\sqrt{s}=200~\mathrm{GeV}$) and the LHC ($\sqrt{s}=13~\mathrm{TeV}$). Figure~\ref{fig:pp_rhic} shows that the dipole amplitude extracted with ciBK evolution reasonably describes the data at both energies. As before, however, different fragmentation-function sets require different $K_h$ factors to match the data. A further notable result is that, for a given FF set, $K_h$ at RHIC is significantly larger than at the LHC, similar conclusion can be found in recent Ref.~\cite{Korcyl:2026nrz}. This trend is consistent with the expectation that higher-order corrections and threshold resummation effects grow more important at lower collision energies. As seen, for instance, in the NLO calculations of Ref.~\cite{Shi:2021hwx}. A separate feature is that the $p_T$-spectrum at $\sqrt{s}=13~\mathrm{TeV}$ is flatter than the data, particularly at high $p_T$. This discrepancy likely calls for resummation of the high-$p_T$ region~\cite{Guzey:2026lfb,Cacciari:2025tgr} or the inclusion of parton-shower effects~\cite{Fujii:2026ccu}.

\section{Factorization scale in PDF and FF}
\label{app:scale}
In this section, we investigate the sensitivity of the nuclear modification factor to the scale choices entering the calculation of Eq.~\eqref{eq:single_hadron} in the main text. In the present leading-order CGC calculation, a common factorization scale $\mu$ is used in both the PDFs and FFs. Following Ref.~\cite{Shi:2021hwx}, the scale is parameterized as
\begin{equation}
    \mu^2
    =
    \alpha^2
    \left(
        \mu_{\rm min}^2+p_T^2
    \right),
    \label{eq:factorization_fragmentation_scale}
\end{equation}
where $p_T$ is the transverse momentum of the final-state hadron and $\mu_{\rm min}=1~\mathrm{GeV}$ sets a lower bound on the scale, preventing the PDFs and FFs from being evaluated at nonperturbatively small scales. To estimate the residual uncertainty associated with this prescription, we compute $R_{p\mathrm{Pb}}$ using the JAM24 FF set~\cite{Anderson:2024evk} while varying the dimensionless scale parameter $\alpha=2,\ 3,\ \text{and}\ 4$, with all other theoretical inputs held fixed. For each value of $\alpha$, the same factorization scale prescription is applied consistently to both the $pp$ and $pA$ cross sections entering the nuclear modification factor.

\begin{figure}[htbp]
    \centering
    \includegraphics[width=0.45\textwidth]{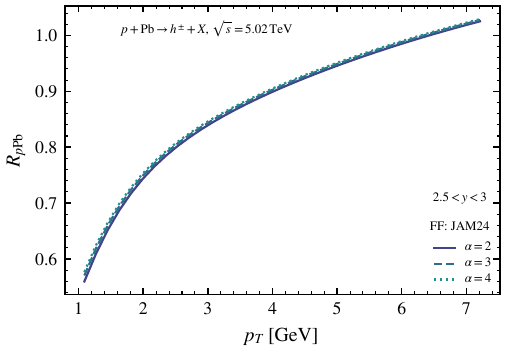}
    \caption{Nuclear modification factor $R_{pA}$ computed using the JAM24 fragmentation functions~\cite{Anderson:2024evk} for three different factorization scales, $\mu^2=\alpha^2(\mu_{\rm min}^2+p_T^2)$ with $\mu_{\rm min}=1~\mathrm{GeV}$, and $\alpha=2$, $3$, $4$. The same scale prescription is applied to the $pp$ and $pA$ cross sections for each value of $\alpha$. The overlap of the three curves shows that $R_{pA}$ is largely insensitive to this scale variation.}
    \label{fig:FF_scale}
\end{figure}

Although the single-inclusive hadron cross sections retain a noticeable dependence on the common scale $\mu$, a substantial part of this correlated dependence cancels in $R_{pA}$. As shown in Fig.~\ref{fig:FF_scale}, the predictions obtained with $\alpha=2$, $3$, and $4$ are nearly indistinguishable over the displayed kinematic range, demonstrating that $R_{pA}$ has only a very weak residual dependence on the scale variations considered here. Together with the reduced sensitivity to the choice of FF sets discussed in previous Sec.~\ref{app:single_inclusive}, this result supports the use of $R_{pA}$ as a robust observable for constraining the initial-state nuclear dipole amplitude and small-$x$ nuclear dynamics.

\section{Exclusive vector-meson production and UPC kinematics}
\label{app:VM}
In the dipole picture, the forward transverse and longitudinal amplitudes are given by
\begin{align}
    \mathcal{A}_{T, L}(x,Q^2) =& \int_0^1 \dd z \int_0^{\infty}r \dd r  (\Psi_V^* \Psi_{\gamma})_{T.L}(r,z,Q^2)N(r,x)
\end{align}
We employ the boosted-Gaussian $J/\psi$ wave function in the calculation~\cite{Kowalski:2006hc},
\begin{align}
    \paren{\Psi_V^* \Psi}_{T} =& \hat{e}_f e {N_c \over \pi z(1-z)}\Big\{ m_f^2 K_0(\varepsilon r)\phi_T(r,z)- [z^2 + (1-z)^2]\varepsilon K_1(\varepsilon r) \partial_r\phi_T(r,z) \Big \},
\end{align}

\begin{align}
    \paren{\Psi_V^* \Psi}_{L} =& \hat{e}_f e{N_c \over \pi} 2 Q z (1-z) K_0(\varepsilon r)\Big[M_V\phi_L(r,z)+ \delta {m_f^2 - \nabla_r^2 \over M_V z (1-z)}\phi_L(r,z) \Big],
\end{align}
where $\hat{e}_f$ is the fractional electric charge of the charm quark, $e=\sqrt{4\pi\alpha_{\text{em}}}$, $m_c=1.4~\mathrm{GeV}$, and  $M_{J/\psi}=3.097~\mathrm{GeV}$. 

The real-part and skewness corrections are computed locally as
\begin{align}
    &\lambda_{T,L} = \partial_Y \ln \mathcal{A}_{T,L}, \quad \beta_{T,L} = \tan{\frac{\pi \lambda_{T,L}}{2}} \notag \\
    &R_g(\lambda) = {2^{2\lambda+3} \over \sqrt{\pi}} {\Gamma(\lambda + 5/2) \over \Gamma(\lambda+4)}
\end{align}
For the impact-parameter-averaged calculation, the coherent cross section is
\begin{equation}
    \sigma_{\gamma A \to VA} = K_V \frac{S_{\perp}^A}{4} \sum_{\lambda=T,L} R_g^2(\lambda)(1+\beta_{\lambda}^2)|\mathcal{A}_{\lambda}|^2,
\end{equation}
where $K_V=1.079$ is fixed from our previous work of $J/\Psi$ photoproductions in $ep$~\cite{Dai:2026nzp}, and $S_{\perp}^A$ is the effective transverse area of the nucleus. The same $S_{\perp}^A$ is used both in $J/\psi$ photoproduction and in forward single-inclusive hadron production in $p\mathrm{Pb}$ collisions. Because the present calculation does not resolve the impact-parameter dependence of the dipole amplitude, it cannot describe the $t$-differential coherent diffractive minima or incoherent contributions; we leave such extensions to future work.

For symmetric Pb--Pb ultraperipheral collisions (UPCs), both photon-emission directions contribute,
\begin{align}
\frac{\dd \sigma_{AA \to AJ/\psi A}}{\dd y} = n(\omega_+)\sigma_{\gamma A}(W_+) + n(\omega_-)\sigma_{\gamma A}(W_-),
\end{align}
where $\omega_{\pm} = \frac{M_{J/\psi}}{2}e^{\pm y}$, $ W_{\pm}^2 = 2\omega_{\pm}\sqrt{s_{NN}}$ is the photon-nucleon center-of-mass energy, and $n(\omega_{\pm})$ is the photon flux, evaluated using the STARlight event generator~\cite{Klein:2016yzr}.

\section{Neural-network implementation and uncertainty propagation}
\label{app:obs_pro}
This Appendix provides the technical details of the neural-network architecture, physics-informed training procedure, loss-weight calibration, and uncertainty propagation underlying the reconstruction presented in the main text.
\subsection{Residual architecture and activation function}
The input $(r,Y)$ is first mapped to an initial hidden representation $\bm h^{(0)}$ by a fully connected embedding layer. It is then processed by a sequence of residual blocks,
\begin{equation}
    \bm h^{(\ell+1)}
    =
    \bm h^{(\ell)}
    +
    W_{2}^{(\ell)}
    \operatorname{Softplus}
    \left(
    W_{1}^{(\ell)}\bm h^{(\ell)}
    +
    \bm b_{1}^{(\ell)}
    \right)
    +
    \bm b_{2}^{(\ell)},
\end{equation}
where $W_{1,2}^{(\ell)}$ and $\bm b_{1,2}^{(\ell)}$ are the weight matrices and bias vectors of the $\ell$-th block. All trainable parameters are randomly initialized and optimized jointly.

We use the Softplus activation, a smooth and differentiable approximation of ReLU, defined as
\begin{align}
    \operatorname{Softplus}(z) = {1\over \beta} \log \left(1 + \exp(\beta z) \right).
\end{align}
The parameter $\beta$ controls the smoothness width. To avoid numerical overflow in the exponential term, the implementation falls back to the linear form $f(z)=z$ when $\beta z$ exceeds a preset threshold. In the present work, we set $\beta=1$ and threshold $=20$. Following the residual blocks, the fully connected layers produce the outputs $g_{\theta}(r,Y)$ and $v_{\theta}(r,Y)$ entering Eqs.~\eqref{eq:NN_amplitude_parameterization} and \eqref{eq:NN_variance_parameterization}.

\subsection{Collocation sampling}
The ciBK residual in Eq.~\eqref{eq:ciBK_loss} is evaluated at $N_{\mathrm{col}}$ collocation points. At each epoch, these points are resampled using a two-dimensional Latin-hypercube design in the transformed domain $(\ln r,Y)$. Sampling in $\ln r$ provides approximately uniform coverage across the ultraviolet, saturation, and large-$r$ regions despite the several-orders-of-magnitude range of the dipole size.

The use of a joint Latin-hypercube design is particularly advantageous for evaluating the nonlocal ciBK residual. A tensor-product discretization based on $N_r$ dipole-size points and $N_Y$ rapidity points would require evaluating the evolution kernel at all $N_rN_Y$ combinations. Since each residual evaluation involves an additional integration over the daughter-dipole size $r_1$ and the relative azimuthal angle, the resulting computational cost scales schematically as
\begin{equation}
    \mathcal O
    \left(
    N_rN_YN_{r_1}N_{\varphi}
    \right).
\end{equation}
By contrast, the two-dimensional Latin-hypercube construction directly generates $N_{\mathrm{col}}$ joint samples $(r_j,Y_j)$, reducing the corresponding cost to
\begin{equation}
    \mathcal O
    \left(
    N_{\mathrm{col}}N_{r_1}N_{\varphi}
    \right),
\end{equation}
with $N_{\mathrm{col}}$ chosen independently of the size of a Cartesian product grid.

Latin-hypercube sampling stratifies each input coordinate and therefore provides more space-filling coverage than an equal number of unstructured random samples. Randomized resampling at every epoch exposes the network to different regions of the $(\ln r,Y)$ domain throughout training. Over sufficiently many epochs, this progressively improves the statistical coverage of the full domain without storing or repeatedly evaluating a prohibitively dense tensor-product grid.

\subsection{Gradient-based loss-weight calibration}
The physics-informed objective combines loss components that encode distinct physical and observational constraints and can exhibit substantially different numerical scales and gradient magnitudes. Without an appropriate relative normalization, one component can dominate the parameter updates while the remaining constraints are only weakly enforced~\cite{doi:10.1137/20M1318043,WANG2024113112}. Motivated by gradient-normalization methods developed in multitask learning and physics-informed neural networks~\cite{DBLP:journals/corr/abs-1711-02257,WANG2024113112}, we determine the relative weights through a preliminary gradient-calibration stage.

We define the loss-component index set as
\begin{equation}
         \mathcal I =\{ \mathrm{ciBK}, R_{pA}, \mathrm{VM}, \mathrm{black}, \mathrm{pos}, \gamma \}, \qquad N_{\mathcal L}=|\mathcal I|=6.
 \end{equation}
 During an initial warm-up stage, all weights are set to unity. At the $t$-th calibration step, the unweighted gradient norm associated with component $a\in\mathcal I$ is
 \begin{equation}
     G_a^{(t)} = \left\| \nabla_{\theta} \mathcal L_a^{(t)} \right\|, \qquad a\in\mathcal I,
     \label{eq:gradient_norm_step}
 \end{equation}
 where $\theta$ denotes the trainable neural-network parameters. To reduce fluctuations associated with the stochastic collocation points and observable batches, the gradient norm is averaged over $T_{\mathrm{cal}}$
 \begin{equation}
     \overline G_a = \frac{1}{T_{\mathrm{cal}}} \sum_{t=1}^{T_{\mathrm{cal}}} G_a^{(t)}.
     \label{eq:average_gradient_norm}
 \end{equation}
 The reference gradient scale is defined as the median over the components in $\mathcal I$,
 \begin{equation}
     G_{\mathrm{ref}}
     =
     \underset{a\in\mathcal I}{\operatorname{median}}
     \left(
     \overline G_a
     \right).
     \label{eq:gradient_reference}
 \end{equation}
 The loss weights used in the subsequent training are assigned directly as
 \begin{equation}
     w_a
     =
     \frac{
     G_{\mathrm{ref}}
     }{
     \overline G_a+\epsilon_G
     },
     \qquad
     a\in\mathcal I,
 \end{equation}
 where $\epsilon_G>0$ prevents numerical instabilities when a gradient norm becomes very small. For $\epsilon_G\ll\overline G_a$, this prescription yields
 \begin{equation}
     w_a\overline G_a
     \simeq
     G_{\mathrm{ref}},
     \qquad
     a\in\mathcal I,
 \end{equation}
 so that the different loss components contribute at comparable gradient scales at the end of the calibration stage.

 Following the calibration stage, the weights $w_a$ are held fixed during subsequent training and are excluded from the learnable parameters. Based on the calibration results, the weights for the various loss terms are manually set to $(w_{\mathrm{ciBK}}, w_{R_{pA}}, w_{\mathrm{VM}}, w_{\mathrm{black}}, w_{\mathrm{pos}}, w_{\gamma}) = (5.0\times10^{5}, 100, 100, 1, 1.0\times10^{4}, 2000)$. These values are adopted uniformly across all subsequent training runs.

\subsection{Observable-level uncertainty propagation}
For compactness, the ensemble-member index is suppressed in the following. For the $i$-th data point in observable class $c\in \{R_{pA},\mathrm{VM} \}$, the theoretical prediction is written as
\begin{equation}
O_{c,i}^{\mathrm{th}}
=
\mathcal F_{c,i}
\left[N_{\theta}\right],
\end{equation}
where $\mathcal F_{c,i}$ contains the Fourier transforms, fragmentation and wave-function convolutions, phase-space integrations, and nonlinear operations appropriate to the observable.

The pointwise variance $\sigma_{N,\theta}^{2}(r,Y)$ associated with the dipole amplitude is propagated through the same forward map. Introducing the collective dipole coordinate $\xi=(r,Y)$, we define the functional Jacobian
\begin{equation}
    \mathcal J_{c,i}(\xi)
    =
    \frac{
    \delta O_{c,i}^{\mathrm{th}}
    }{
    \delta N_{\theta}(\xi)
    }.
\end{equation}
To first order in the fluctuations of the dipole amplitude, the neural-network contribution to the observable covariance is
\begin{equation}
    C_{c,ij}^{\mathrm{NN}}
    =
    \int \dd\xi\,\dd\xi'\,
    \mathcal J_{c,i}(\xi)\,
    C_N(\xi,\xi')\,
    \mathcal J_{c,j}(\xi'),
    \label{eq:obs_covariance_continuous}
\end{equation}
where
\begin{equation}
    C_N(\xi,\xi')
    =
    \operatorname{Cov}
    \left[
    N_{\theta}(\xi),
    N_{\theta}(\xi')
    \right]
\end{equation}
is the covariance of the dipole amplitude.

In the implementation, the forward maps are discretized at the quadrature nodes $\xi_a=(r_a,Y_a)$. We approximate the covariance of the network outputs as diagonal in this discretized representation,
\begin{equation}
    C_{N,ab}
    \simeq
    \delta_{ab}\,
    \sigma_{N,\theta}^{2}(\xi_a),
    \label{eq:dipole_covariance_diagonal}
\end{equation}
thereby neglecting correlations between different dipole-input points. Defining the Jacobian of the discretized forward map as
\begin{equation}
    J_{c,ia}^{\mathrm{disc}}
    =
    \frac{
    \partial O_{c,i}^{\mathrm{th}}
    }{
    \partial N_{\theta}(\xi_a)
    },
\end{equation}
the propagated covariance becomes
\begin{equation}
    C_{c,ij}^{\mathrm{NN}}
    \simeq
    \sum_a
    J_{c,ia}^{\mathrm{disc}}\,
    J_{c,ja}^{\mathrm{disc}}\,
    \sigma_{N,\theta}^{2}(\xi_a).
    \label{eq:obs_covariance_discrete}
\end{equation}
Because $J_{c,ia}^{\mathrm{disc}}$ is evaluated by automatic differentiation through the discretized forward map, it already contains the quadrature weights and integration measures. This procedure includes the nonlinear ratio defining $R_{pA}$ and the squared production amplitude entering $\sigma^{\mathrm{VM}}$.

For the training likelihood, we neglect correlations between distinct data points and retain only the diagonal observable variance,
\begin{equation}
    \left(
    \sigma_{c,i}^{\mathrm{NN}}
    \right)^2
    =
    C_{c,ii}^{\mathrm{NN}}
    \simeq
    \sum_a
    \left(
    J_{c,ia}^{\mathrm{disc}}
    \right)^2
    \sigma_{N,\theta}^{2}(\xi_a).
    \label{eq:obs_variance_propagation}
\end{equation}
Assuming that the propagated neural-network uncertainty and the experimental uncertainty are independent, the total variance is
\begin{equation}
    V_{c,i}
    =
    \left(
    \sigma_{c,i}^{\mathrm{NN}}
    \right)^2
    +
    \left(
    \sigma_{c,i}^{\mathrm{exp}}
    \right)^2.
\end{equation}
The observable loss is therefore
\begin{align}
    \mathcal L_{\mathrm{obs}}
    =
    \sum_c
    \frac{1}{N_c}
    \sum_{i=1}^{N_c}
    \left[
    \frac{
    \left(
    O_{c,i}^{\mathrm{th}}
    -
    O_{c,i}^{\mathrm{exp}}
    \right)^2
    }{
    2V_{c,i}
    }
    +
    \frac{1}{2}
    \ln V_{c,i}
    \right],
\end{align}
where $N_c$ is the number of data points in observable class $c$. The logarithmic term penalizes overly large learned uncertainties and prevents the network from trivially reducing the standardized residual by inflating $V_{c,i}$.

\clearpage
\twocolumngrid

\bibliography{pA_refs}

\end{document}